\documentclass[12pt]{iopart}

\usepackage{iopams}
\usepackage{graphicx}
\begin{document}

\title[How to make large, void-free dust clusters in dusty plasma under microgravity.]{How to make large, void-free dust clusters in dusty plasma under microgravity.}

\author{V Land $^1$ and W J Goedheer $^2$}

\address{$^1$ Center for Astrophysics, Space Physics, and Engineering Research, Baylor University, Waco, TX, USA 76798-7316\\
$^2$ FOM-Institute for plasma physics `Rijnhuizen', Edisonbaan 14, 3430 BE Nieuwegein, the Netherlands, www.rijnhuizen.nl}
\ead{victor\_land@baylor.edu}

\begin{abstract}
Collections of micrometer sized solid particles immersed in plamsa are used to mimic many systems from solid state and fluid physics, due to their strong electrostatic interaction, their large inertia, and the fact that they are large enough to be visualized with ordinary optics. On Earth, gravity restricts the so called dusty plasma systems to thin, two-dimensional layers, unless special experimental geometries are used, involving heated or cooled electrons, and/or the use of dielectric materials. In micro-gravity experiments, the formation of a dust-free void breaks the isotropy of three-dimensional dusty plasma systems. In order to do real three-dimensional experiments, this void has somehow to be closed. In this paper, we use a fully self-consistent fluid model to study the closure of a void in a micro-gravity experiment, by lowering the driving potential. The analysis goes beyond the simple description of the ''virtual void'', which describes the formation of a void without taking the dust into account. We show that self-organization plays an important role in void formation and void closure, which also allows a reversed scheme, where a discharge is run at low driving potentials and small batches of dust are added. No hysteresis is found this way. Finally, we compare our results to recent experiments and find good agreement, but only when we do not take charge-exchange collisions into account.

\end{abstract}


\maketitle

\section{Introduction: Dusty plasma as a model system}
The difference between dusty plasma and ordinary plasma is the presence of small solid particles that collect electrons and ions from the surrounding plasma. This means that the dust particles charge up, and many forces start acting on them. The electrostatic interaction between the shielded dust particles determines the internal structure of the dust clouds. The coupling between the particles depends on the ratio of their mutual Coulomb interaction potential and their kinetic energy, which is usually referred to as the ''coupling parameter'', $\Gamma = Q_d^2\exp(-\Delta/{\lambda}_d)/\Delta k_bT_d$. $Q_d$ is the dust charge, $\Delta = (3/4\pi n_d)^{1/3}$, the inter-particle distance, and $T_d$ the dust kinetic temperature. It is worthwhile to mention that the latter is very hard to determine, but different methods are used, see for instance \cite{WilliamsPOP2007}. 

When ${\Gamma \gg 1}$ crystalline structures form, while for ${\Gamma \ll 1}$ the dust particles show fluid-like or gas-like behavior. Such dusty plasma systems are ideally suited to study, for instance, waves and phonons in solids, or phase-transitions, but also the laminar and turbulent flow properties of a liquid \cite{LeePRE1997,NunomuraPRL2002,MorfillPRL2004}, because the dust particles are big enough to be visualized with ordinary optical techniques. Big particles also imply that in experiments on Earth gravity is the dominant force. This confines the particles to thin, two-dimensional layers. A thermophoretic force can be added to balance the force of gravity by heating, or cooling the electrodes in the experiments, which is usually only done for small dust clusters containing a few thousands of particles or less \cite{BlockPPCF2007}, however, an example of a similar experiment with a large number of particles can be found in \cite{Rothermel2002}.

In microgravity experiments performed on the International Space Station, large, three-dimensional dust structures were obtained. However, instead of isotropic dust crystals, a large dust-free void was found, resulting in a large anisotropy \cite{MorfillPRL1999}. This void results from the interaction between ions, which move from the center of the discharge to the outer walls, and the dust particles, and is a perfect example of self-organization in dusty plasma \cite{LandNJOP2006,LandNJOP2007}. 

In order to create large, isotropic, three-dimensional dust structures, the void somehow has to be closed. One way to achieve this is by reducing the driving potential, so that the ion density decreases. One has to be careful not to terminate the discharge this way. In a recent experiment \cite{LipaevPRL2007} exactly this was done. The authors reported the closure of the void and analyzed the results using the `virtual void' formalism. This is a description of the void size in terms of the balance between the inward electrostatic force and the outward ion drag force, \emph{without} taking the effect of the dust on the plasma parameters into account. It is therefore valid only to describe experiments with a very limited number of dust particles, such as in \cite{KlindworthJOPD2006}. This was however not the case in their experiment.  

In this paper we perform numerical experiments to study the void closure in dusty plasma under microgravity, trying to bring more insight in the physics behind void closure in experiments, performed both in the recent past, as well as in current state-of-the-art machines \cite{ThomasNJOP2008}. The modelled discharge settings are exactly equal to those in \cite{LipaevPRL2007}, so that we can directly compare our results and analysis with those presented there. We show that the self-organization in dusty plasma plays an essential role. Furthermore, we study the effect of charge-exchange (cx-) collisions on dust charging and void closure, as these collisions are expected to be important in the type of dusty plasmas considered.

\section{Numerical tool}

A complete description of our model is given in \cite{LandNJOP2006}, here we restrict ourselves to a shorter description. Our model solves both the plasma parameters, as well as the dust parameters in a coupled fashion. We start by briefly describing the solution for the plasma parameters.

\subsection{The plasma parameters}
Our fluid model solves the balance equation for a quantity $A$:

\begin{equation}\label{quantity-balance}
\frac{\partial A}{\partial t} = -\nabla\cdot\boldsymbol{\Gamma}_A + S_A,
\end{equation}

\noindent assuming a drift-diffusion expression for the flux ${\Gamma}_{A}$:

\begin{equation}
{\boldsymbol{\Gamma}}_{A} = {\mu}_{A} A \boldsymbol{E} - D_{A}\nabla A,
\end{equation}

\noindent which is then used in equation \ref{quantity-balance}. Here $S_A$ represents the sources and sinks of quantity $A$, ${\mu}_A$ the mobility, and $D_{A}$ the diffusion coefficient. We solve these equations for $A=n_{e,+}$, for which we have the mobility as ${\mu}_{e,+} = \mp e/m_{e,+}{\nu}_{e,+}$ with ${\nu}_{e,+}$ the electron (ion) momentum transfer frequency. The diffusion coefficient is given by the Einstein relation $D_{e,+}=k_bT_{e,+}{\mu}_{e,+}$. The source/sink terms in this case include electron impact excitation, 
ionization, and recombination on the dust particles. For $A=w_e=n_e\epsilon$, we have ${\mu}_{w_e} = \frac{5}{3}{\mu}_e$. The sources and sinks are the same, but now also include the Ohmic heating term as a source, $S_{Ohmic}=\boldsymbol{J_e}\cdot\boldsymbol{E}$. The ions are assumed to dissipate their heat instantaneously, so that they are in equilibrium with the neutrals. Therefore, we do not solve a similar equation for the ion energy density.

In order to solve the above equation, we need to find the electric field from the Poisson equation:

\begin{equation}\label{Poisson}
-{\nabla}^2 V = -\frac{e}{{\epsilon}_0}\left(n_e+n_dZ_d-n_{+}\right), ~~\boldsymbol{E}=-\nabla V,
\end{equation}

\noindent with $Z_d$ the dust charge number. The ions are too heavy to follow the instantaneous field $\boldsymbol{E}$, but instead an effectieve field is found by solving $d {\textbf{E}}_{eff}/dt = {\nu}_{+}(\boldsymbol{E}-{\boldsymbol{E}}_{eff})$.

All these equations are solved on sub-RF time-steps. However, the presence of $Z_d$ in the Poisson equation, as well the source terms for electron-ion recombination on the dust requires the solution of the dust density and charge.

\subsection{The dust parameters}
The calculation of the dust charge is done by solving the current of electrons and ions to the dust particles using Orbital Motion Limited theory \cite{AllenPS1992}, with the local plasma parameters as input. In equilibrium, the currents balance, $I_{+} + I_e = 0$, and the surface potential $V_d$ is found. Assuming that the dust particles act as capacitors, the dust charge (number) is then found from $Q_d = eZ_d = 4\pi {\epsilon}_0 r_d V_d$. In case we include charge exchance collisions, the ion current towards the dust particles is higher than the OML current. We calculate it with the analytic method derived in \cite{LampePRL2001}.

The dust transport is solved by calculating the electrostatic force, ${\textbf{F}}_{\overline{E}}$, the ion drag force, ${\textbf{F}}_{ion}$, the thermophoretic force, ${\textbf{F}}_{th}$, and the dust diffusion, which depends on the equation of state of the dust \cite{GozadinosNJOP2003}. These forces are assumed to be in balance with the neutral drag ${\textbf{F}}_{nd} = m_d{\nu}_{m,d}{\boldsymbol{\Gamma}}_d/n_d$, so that the dust flux can be written as:

\begin{equation}\label{dustflux}
{\boldsymbol{\Gamma}}_d = \frac{n_d{\textbf{F}}_{\overline{E}}}{m_d{\nu}_{m,d}} - D_d\nabla n_d + \frac{n_d{\textbf{F}}_{th}}{m_d{\nu}_{m,d}} + \frac{n_d{\textbf{F}}_{ion}}{m_d{\nu}_{m,d}},
\end{equation}

\noindent where ${\nu}_{m,d} = (4\pi r_d^2/3)(n_{gas}v_T(m_n/m_d))$, is the dust-neutral momentum transfer frequency, and $\overline{E}$ is the time averaged (over a RF period) electric field. The ion drag force includes the effect of moderate non-linear scattering, anisotropic screening due to ion drift, as well as the effect of cx-collisions \cite{KhrapakPRE2002,HutchinsonPPCF2006,IvlevPRL2004}. The latter is similar to the effect of these on the dust charging; a hot ion ''collides'' with a cold atom, losing angular momentum and energy, so that the probability to be captured by a nearby dust particle is increased. This increases the net ion current to the dust particles, increasing both the ion drag force \emph{and} reducing the negative charge on the dust. This in turn has the secondary effect of reducing the ion drag force, since this force is $\propto Z_d^2$, as will be shown later. The effect of these collisions on the ion drag force is always taken into account. We distinguish between cases with and without the effect on the charging.

The dust transport is solved on longer time-steps. To maintain quasi-neutrality, the ion density is adapted, which causes a growing defect in the solution of the plasma densities. When this defect becomes too large, the dust is frozen and the plasma parameters are calculated again on sub-RF time-steps. The plasma and dust parameters are then completely coupled via the Poisson equation and the source-terms and we compute towards the final equilibrium solution. Extending the dust flux equation to a complete momentum equation, by including the total derivative of the density, to be able to look at the dynamic behavior of the dust transport, is something for future work, and is not included here.

\section{The modelled discharge}

The experiment we model is the Plasma Kristall Experiment, which was placed on board of the International Space Station. This is a cylindrically symmetric RF discharge, run with argon gas. We use the following discharge parameters: neutral pressure, $P_{gas}$ = 24 Pa, background gas temperature, $T_{gas} = 293 K$, RF-frequency ${\nu}_{RF}$ = 13.56 MHz. We simulate melamine formaldehyde dust particles with a diameter of 6.8 micron. Because of the symmetry, we only model half of the discharge in the r-z plane. The determination of the void size for the different values of the driving potential ($V_{RF}$, peak-to-peak potential $V_{pp} = 2V_{RF}$) is explained below.

\begin{figure}[h]
\center
\includegraphics[width=0.6\textwidth]{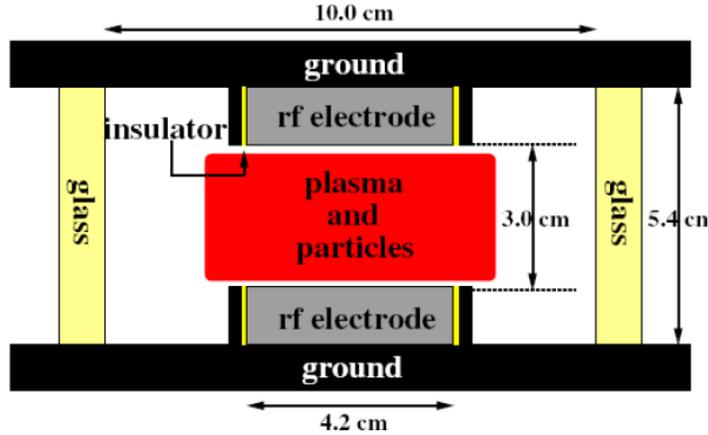}
\caption{The geometry of the experiment we model (the Plasma Kristall Experiment, PKE).}\label{fig:geometry}
\end{figure}

\section{Determination of void size}

\subsection{Virtual void size}
To follow the analysis in \cite{LipaevPRL2007}, we first determine the \emph{virtual} void size in the discharge. Dust particles added to a discharge will move to points where the total force acting on them vanishes. When only a small number of particles is added to the discharge, so that the plasma parameters are not influenced by the losses on the dust, these points form a contour where the outward ion drag force balances the inward electrostatic force, since the thermophoretic force is much smaller. This contour is called the virtual void. We determine the virtual void size, $z_v$, as the distance between the central point and the point where this contour crosses the axial symmetry line, in simulations where we do not add any dust particles. So, we calculate the forces that would act on a dust particle, if it would have been added to the discharge, and from that find the point on the symmetry axis where this force vanishes, which is a rather straightforward method.

\subsection{Real void size}
The determination of the \emph{real} void size, $z_r$, when a large number of dust particles is added is less straightforward. The contour where the ion drag and electrostatic force balance no longer coincides with the real inner boundary of the void, due to the pressure inside the dust cloud. In the experiment, an estimate for the dust density is made from the inter-particle distance. From this then, the inner boundary of the dust cloud can also be determined. We determine the real void size as that point on the symmetry axis where the dust density drops below $ n_c = {10}^9$ m${}^{-3}$. This choice is arbitrary, but corresponds roughly to 10 \% of the maximum density. 

Of course, it is rare that the density on one of the grid points in our simulation has exactly this value. Therefore, we need to interpolate the dust density between grid points. We designate the real void size to be the distance between the point where the interpolated line (or `fit') then crosses this value and the center of the discharge, again, along the symmetry axis. The interpolation we use is not arbitrary, but based on the exponential scheme used to solve the dust flux. This exponential scheme is based on the assumption of a constant drift-diffusion flux between successive grid points.

In one direction (along the symmetry axis), this reads:

\begin{equation}
{\Gamma}_d = n_d u_d - D_d{\partial}_zn_d = C, 
\end{equation}

\noindent with $u_d$ the dust speed from the forces acting on the dust. Trying a solution of the form $n_d(z) = A + B ~\exp(\alpha z)$, with 2 boundary conditions for $n_d(z)$, the lower value $n_d(z_1) = n_l < n_c$, and the upper value $n_d(z_2) = n_u > n_c$, where $z_2 = z_1 + \Delta$, with $\Delta$ the grid-interval, we find the solution as,

\begin{equation}
n_d(z) = n_l + \frac{n_u-n_l}{\exp\left(\frac{u_d}{D_d}\Delta\right)-1}\left(\exp\left(\frac{u_d}{D_d}(z-z_1)\right)-1\right), ~~ z\in(z_1,z_2).
\end{equation}

\noindent To find $z_r$, we need to solve

\begin{equation}
n_d(z_r) \equiv n_c = n_l + \frac{n_u-n_l}{\exp\left(\frac{u_d}{D_d}\Delta\right)-1}\left(\exp\left(\frac{u_d}{D_d}(z_r-z_1)\right)-1\right), ~~ z_r\in(z_1,z_2).
\end{equation}

\noindent This way, we find

\begin{equation}
z_r = z_1 + \frac{D_d}{u_d}\ln\left[1+\frac{n_c-n_l}{n_u-n_l}\left(\exp\left(\frac{u_d}{D_d}\Delta\right)-1\right)\right].
\end{equation}

\section{Results: Virtual void sizes}

\begin{figure}[h]
\includegraphics[width=0.45\textwidth]{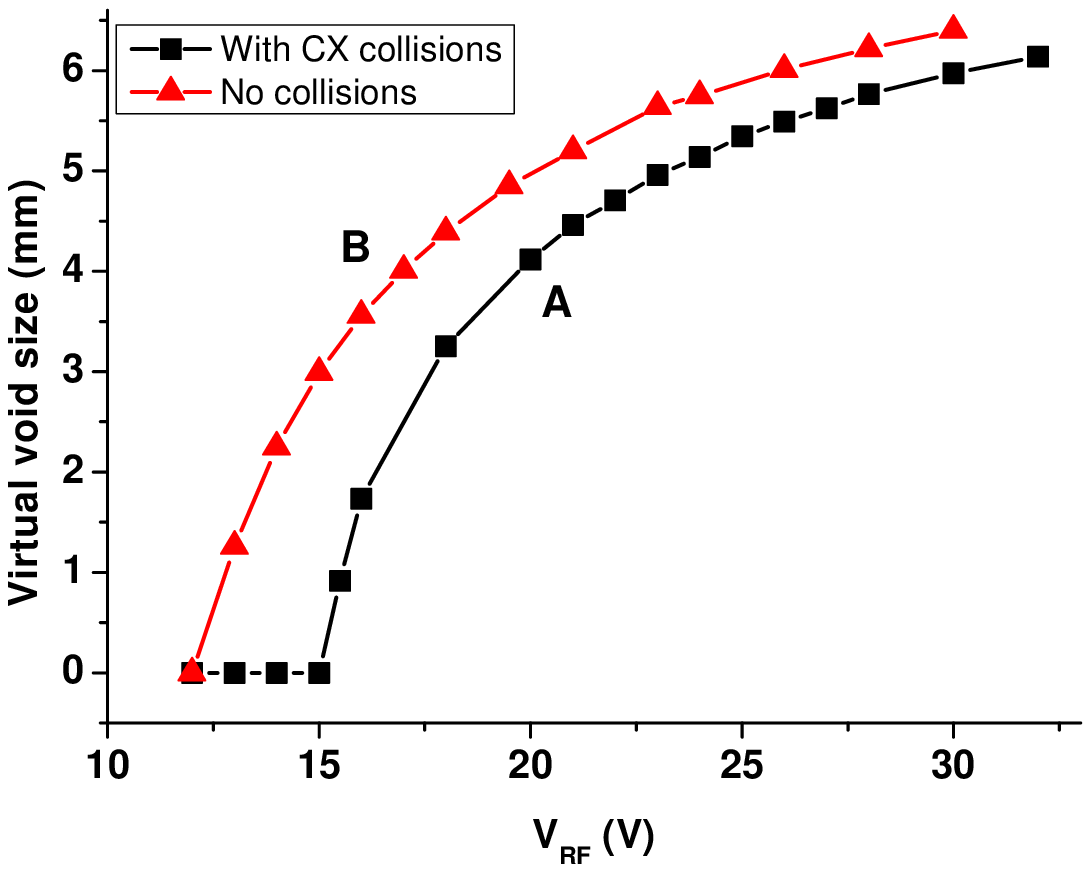}
\includegraphics[width=0.45\textwidth]{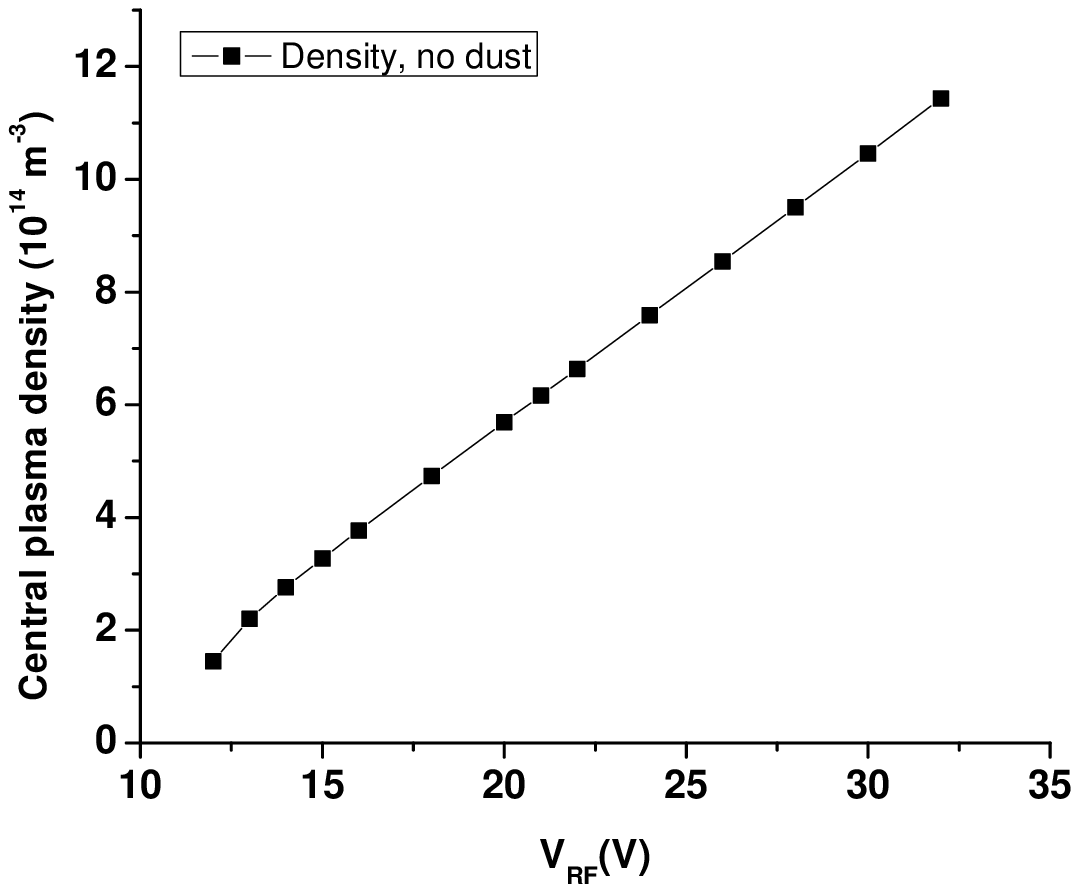}
\caption{\textbf{Left:} The obtained virtual void sizes with cx-collisions, indicated by the black squares and line, labeled 'A', and without cx collisions, indicated by the red triangles and line, labeled 'B'. \textbf{Right:} The central electron density without any dust present (because of quasi-neutrality, it equals the central ion density), for different values of the driving potential. Clearly, the plasma density drops linearly for decreasing driving potential.}\label{fig:virtualvoidsize}
\end{figure}

We start with a dust free discharge, running at $V_{RF} \sim$ 30 V. We determine the virtual void size, and then reduce the potential in small steps. At different values of the driving potential we calculate the virtual void size. We do this set of calculations with and without cx-collisions (which is from now on to be understood as; with and without the effect of cx-collisions on the dust charge). The results are shown in figure \ref{fig:virtualvoidsize} on the left.

First of all, we indeed see that the void decreases for decreasing driving potential, which coincides with the decrease in plasma density, as shown in figure \ref{fig:virtualvoidsize} on the right. A decrease in ion density results in a decrease of the ion drag force, since the collected and scattered ion flux is proportional to the ion density.

Secondly, we see that the void sizes with collisions are simply shifted downwards with respect to the solutions without collisions. This is not due to any changes in the plasma because of additional plasma recombination, since in these virtual void calculations, no dust is added to the simulation. However, the dust charge is evaluated everywhere, with the plasma parameters as input, even without dust particles present. With collisions, the calculated ion current towards dust particles increases, so that the dust charge becomes less negative. Since the ion drag force $\propto Z_d^2$, and the electrostatic force $\propto Z_d$, the virtual void size decreases when the charge decreases when the collisions are taken into account.

Thirdly, we see that for larger driving potentials the two solutions approach each other, whereas for lower driving potentials there is a large difference between the two. For increasing driving potential, the plasma density increases, so that the Debye length decreases. This means that on average a fast ion approaching a dust particle collides less times within a Debye length. Thus, the ion is unable to lose sufficient energy and angular momentum to reach the dust particle. The additional ion current towards the dust because of cx-collisions is therefore less, and the dust charge will be closer to the value without any cx-collisions.

Finally, we see that the virtual void doesn't decrease exactly proportional to the plasma density. For high potential, the decrease is linear, but for lower driving potential, approaching the point of vanishing virtual void, the decrease becomes much steeper. The virtual void size is determined by the ratio of the outward ion drag force over the inward electrostatic force, $\eta=F_{ion}/F_E$, since the virtual void is that point where this ratio is \emph{exactly} one. Writing out the two forces (using the scattering part of the ion drag only, which is much larger than the collection part), this ratio becomes:

\begin{equation}\label{eq:forceratio}
\eta (z)= \left[\frac{e^3Z_d(z){\mu}_{+}}{4\pi{\epsilon}_0^2m_{+}}\right]\times\frac{n_{+}(z)}{v_s^3(z)} \approx 6\cdot{10}^{-7}\frac{n_{+}(z)}{v_s^3(z)},
\end{equation}

\noindent where $v_s = \sqrt{v_{T+}^2+u_{+}^2}$ is the total ion velocity, $v_{T+}=\sqrt{T_{+}/m_{+}}$ the ion thermal velocity and $u_{+} = {\mu}_+E$ the ion drift velocity. Assuming a constant dust charge, $Z_d \approx {10}^4$, we see that for typical plasma densities ($n_{+} \approx {10}^{15}$ m${}^{-3}$) the total ion velocity at the virtual void edge ($\eta \equiv 1$) should be (2-3) $\times v_{T+}$, consistent with previous simulations \cite{LandNJOP2006}.

The virtual void size is thus determined by the density profile as well as by the drift velocity of the ions. In terms of the drift velocity, there are two limits for equation \ref{eq:forceratio}, the first when the ion drift is subthermal ($u_{+}^2 \ll v_{T+}^2$), and second when the drift is suprathermal ($u_{+}^2 \gg v_{T+}^2$). In the first limit ${\eta} \propto n_{+}/v_{T+}^3$, so, $\eta \propto n_{+}$ (since $v_{T+}$ is constant). The latter limit gives ${\eta} \propto n_{+}/u_{+}^{3}$. One could argue that the decrease in $n_{+}$, and the corresponding increase in the Debye length, results in an increase in $u_{+}$, because of an increase in charge separation and the corresponding electric field. Then, the change in the virtual void size curve could be due to this transition from subthermal ion flow to suprathermal ion flow. Even though this might play a role, the basic shape of the curve can be completely explained by a decrease in density only.

\begin{figure}[h]
\includegraphics[width=0.55\textwidth]{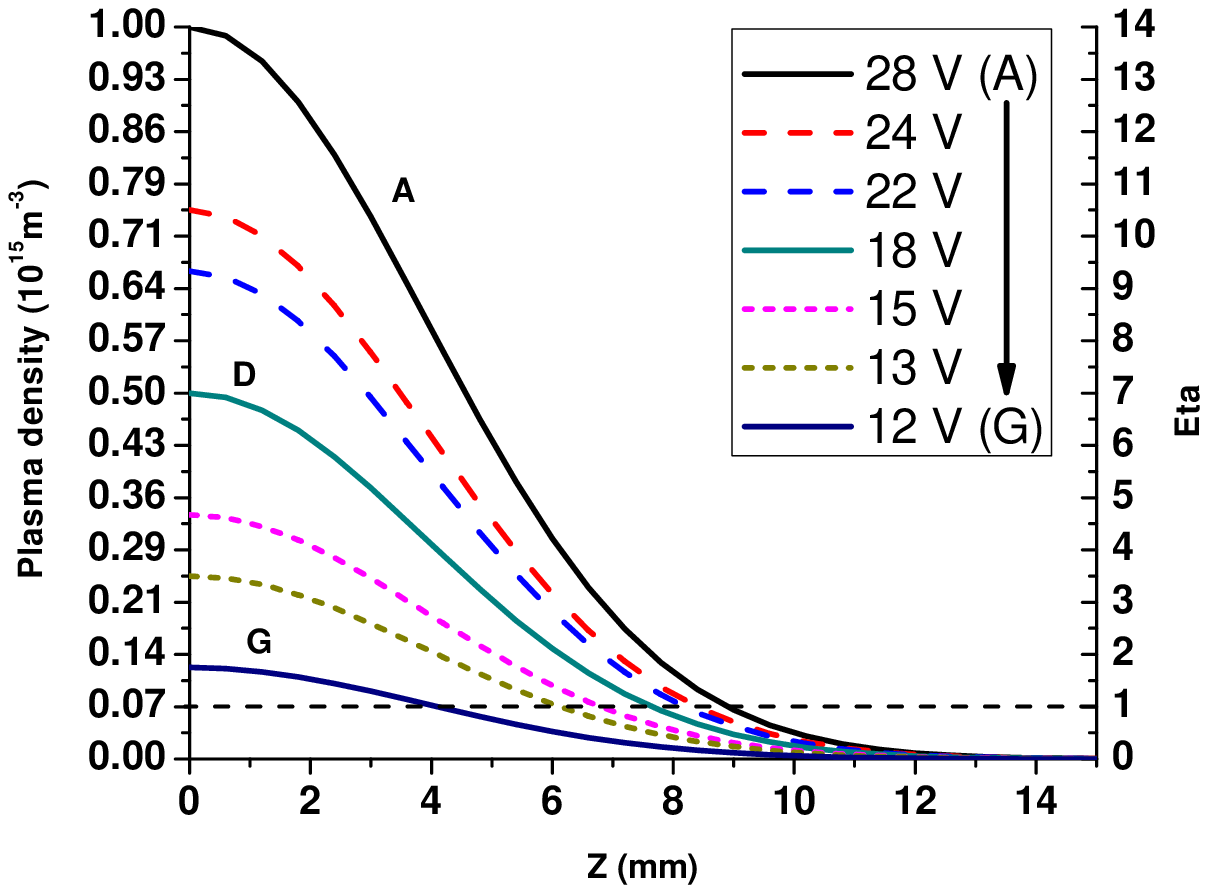}
\includegraphics[width=0.45\textwidth]{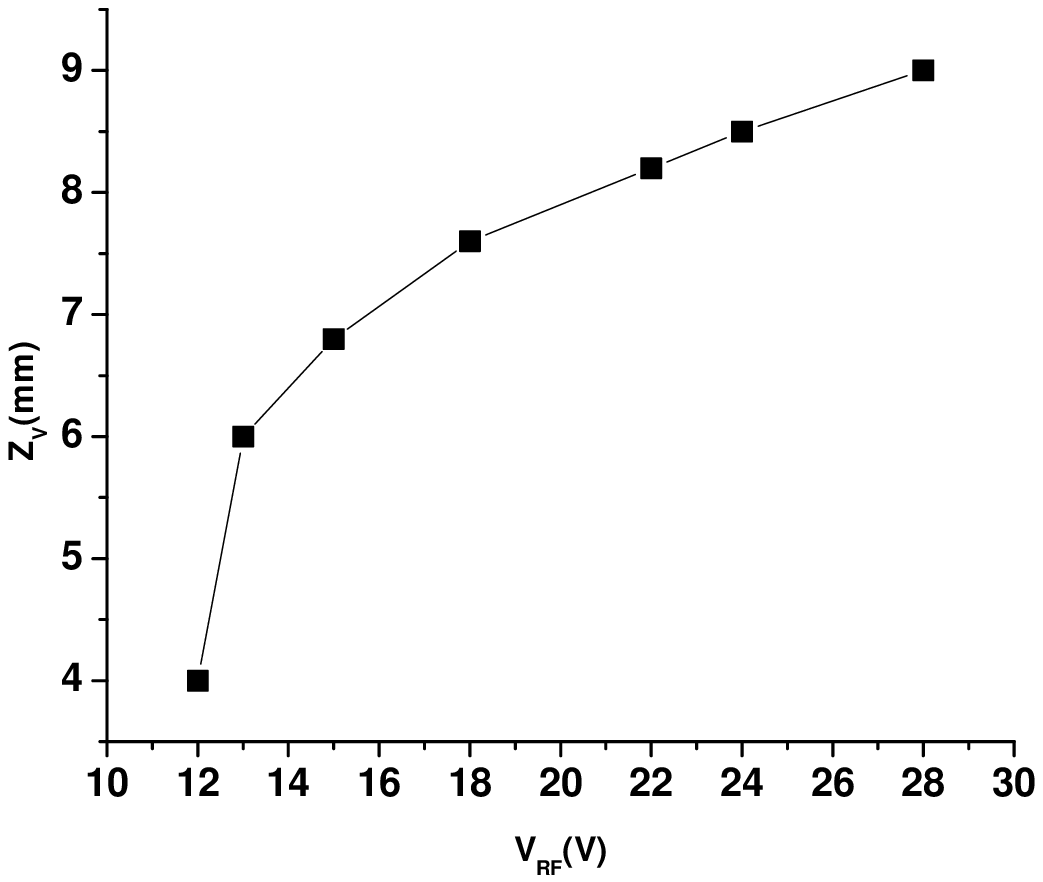}
\caption{\textbf{Left:} We approximate the axial density with an exponentially decaying distribution. The central density at $V_{RF}=28V$ sets ${\eta}_{<} \sim 14$ there. The density vanishes at the upper electrode at z=15mm. For decreasing $V_{RF}$ the density profile flattens. \textbf{Right:} The virtual void size, determined from the points where $\eta=1$ for the different driving potentials in the figure on the left. Even though the exact values are not right, the shape of the virtual void curve found by our simulations is well represented.}\label{fig:uitlegeta}
\end{figure} 

The effect of the changing density is illustrated in figure \ref{fig:uitlegeta} with a simple qualitative model, for subthermal ion drift, so that $\eta \propto n$. For illustrative purposes, we have assumed that the density along the symmetry axis can be described by a bell-shaped curve. The boundary values are the central value of the density on the left, given by figure \ref{fig:virtualvoidsize}, and vanishing density at the upper electrode at z=15 mm. For decreasing driving potential, the central density decreases and this leads to a more and more flat density profile. The central value for $V_{RF}=28V$, $\sim{10}^{15}$ m${}^{-3}$, gives ${\eta} \sim 14$. which fixes the right axis. The flatter the density profile, the larger the virtual void has to shift inwards in order to reach the point $\eta=1$. Plotting the virtual void size found this way against $V_{RF}$ gives the qualitative picture on the right. We see that, even though the exact numbers are not correct, the shape of the virtual void as found in figure \ref{fig:virtualvoidsize} is well reproduced. This shows that the inward shift of the virtual void with decreasing potential is due to the decrease in ion density inside the virtual void.

Now that we have an understanding of the behavior of the virtual void in a plasma without any dust particles added, so that we understand where a very small number of dust particles would reside at different driving potentials, we continue by investigating the behavior of the real void edge, when hundreds of thousands of dust particles are added and the plasma parameters change dramatically.

\section{Results: Real void sizes}

\subsection{The initial dusty plasma}

\begin{figure}[h]
\center
\includegraphics[width=0.8\textwidth]{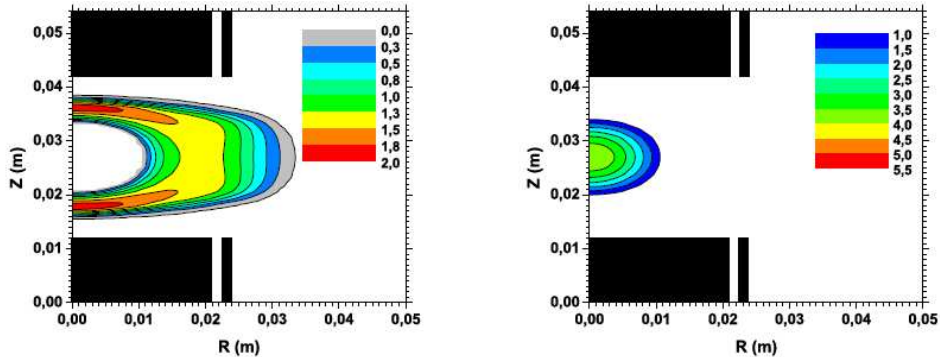}
\caption{The initial dust density in ${10}^{10}$ m${}^{-3}$ on the left, and the electron density on the right in ${10}^{15}$ m${}^{-3}$ for a dust cloud containing 500.000 particles. The void is clearly seen, and also that the plasma is confined within the volume of the void.}\label{fig:dustyplasmabegin}
\end{figure}

A typical solution for the dusty plasma with void is shown in figure \ref{fig:dustyplasmabegin}. The void is clearly seen, and the plasma is confined inside the void. The dust density is of the order ${10}^{10}$ m${}^{-3}$, and the electron density is of the order of ${10}^{15}$ m${}^{-3}$. The electron temperature $T_e$ is roughly 3 $\sim$ 4 eV. By eye, the dust size would be roughly between 6 and 7 mm.

\begin{figure}[ht]
\center
\includegraphics[width=0.5\textwidth]{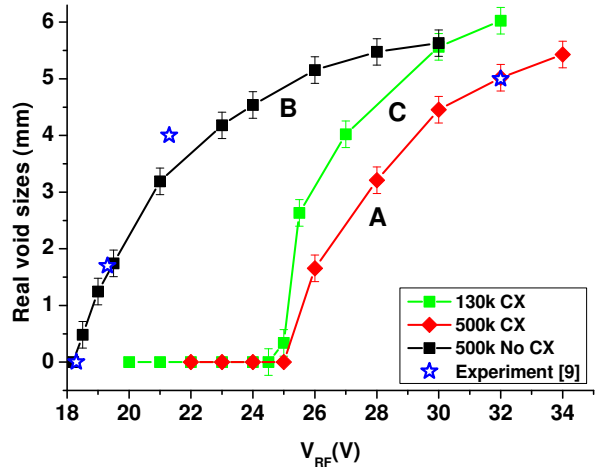}
\caption{The real void sizes for different simulations. The black squares indicate real void sizes for a total of $5\cdot{10}^5$ dust particles added to the discharge, without taking cx-collisions into account. The red diamonds indicate the same number of dust particles, but now with cx-collisions. The blue stars are experimental results from \cite{LipaevPRL2007}. The blue star with $z_r = 0$ is not plotted in their results but mentioned in the text. The green squares is a run with $1.3\cdot{10}^5$ particles, with cx-collisions. The error bars indicate a standard error due to the exponential fit to the density.}\label{fig:realvoidsizes}
\end{figure}

\noindent The real void sizes for different simulations are shown in figure \ref{fig:realvoidsizes}. Starting with the simulations with $5\cdot{10}^5$ particles with (red line and symbols, labeled 'A') and without (black line and symbols, labeled 'B') cx-collisions, we see that they show the same behavior as the virtual void solutions. However, the potentials at which the real voids close are much higher. This is due to the additional losses of plasma on the dust particles and consequently the larger decrease in the ion density and ion drag force. We also see that the solutions are further apart in these results with dust particles. Not only is the charge on the dust particles reduced by the enhanced ion current, but by the same mechanism the recombination rate of plasma on the dust is even more increased, so that the solutions are much further apart, the void without collisions closing at $V_{RF}=18.2 V$, the solution with collisions at $V_{RF}=25V$. 

When the voltage is decreased below the value needed to sustain the discharge, the ionization is no longer sufficient to compensate the losses at the walls and on the dust. Then, the electron density goes to zero, and the remaining dust-ion plasma will decay by recombination and slow ambipolar diffusion, eventually leading to shut-down of the discharge. The final point without collisions is very close to this potential, so that we could not go much lower in driving potential while keeping a stable void-free dusty plasma. The solution with collisions, however, was much more stable and did allow a significant range of driving potentials, while at the same time maintaining a void-free dusty plasma, as indicated by the points all having $z_r=0$. The same holds for the solution with 130.000 particles added (green line and symbols, labeled 'C'). This means that the same solution can also be found by starting with a dust-free discharge at low potential and by adding small numbers of dust, while increasing the potential. This scheme might be less prone to shut-down of the plasma and might provide a more suitable route to obtaining void-free dust crystals. 

\begin{figure}[ht]
\center
\includegraphics[width=0.6\textwidth]{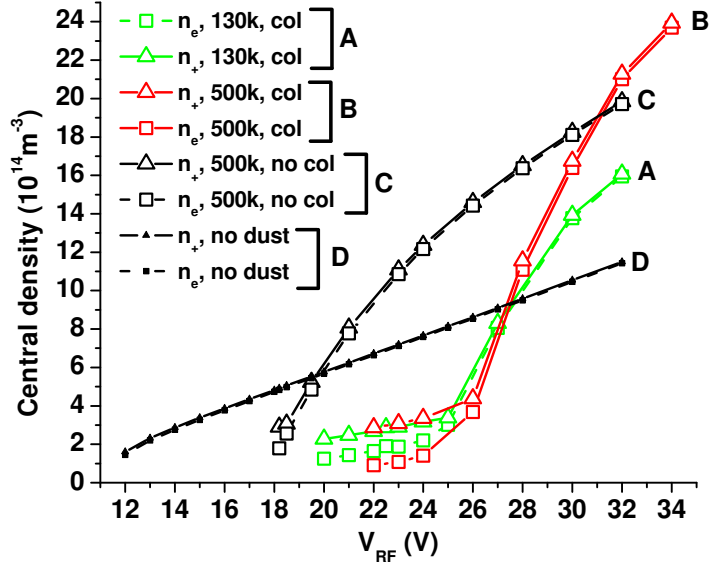}
\caption{The central density for the different runs. The squares represent the electron density, the triangles the ion density. The colors represent different sets of runs, the green, labeled 'A' has 130.000 particles, with cx-collisions, the red, labeled 'B' has 500.000 particles, with cx-collisions, the black has 500.000 particles, without collisions. The densities without dust are included for comparison, shown by the straight black lines with small symbols, labeled 'D'.}\label{fig:densitiesinsidevoid}
\end{figure}

It is interesting to note that the void size for higher driving potentials becomes larger than in the case with 500.000 particles, without collisions. When we look at the electron and ion densities inside the void, shown in figure \ref{fig:densitiesinsidevoid}, we see the effect of the dust on the plasma parameters; the plasma density is increased inside the void, as long as the driving potential is high enough. This was also discussed in \cite{LandNJOP2007}. The additional losses are compensated by additional ionization, for which electron heating due to the increased resistivity provides the necessary energy. The void shows self-organization, through which it maintains itself, despite the plasma-losses. However, once the driving potential becomes too low, the system can no longer provide enough energy to maintain the additional ionization, and the self-organization is lost, so that the plasma density decreases much faster inside the void, until it closes.

A charge-separation arises once the void closes, especially in the case of cx-collisions, where the plasma absorption is very large. Effectively, the resulting dusty plasma consists mainly of negatively charged dust particles and free ions. After void closure, the net space charge will be zero, but the moments before the closure of the void, there is a large positive charge inside the void, resulting in an electric field much larger than the ambipolar field, so that ion drifts are significant.

Returning to figure \ref{fig:realvoidsizes}, we have added some of the experimental points presented in \cite{LipaevPRL2007}. In that experiment several millions of particles were present in the discharge, but from the figures it was clear that not all the particles were involved in the closing of the void, but only a small subgroup of the particles in the cloud, close to the center of the discharge. In our simulations we find a good agreement of our simulations with 500.000 particles with the experimental points without cx-collisions, especially at the lower driving potentials, near void-closure. The point of void-closure reported was 18.3 Volts, even though this point is not depicted in a graph. It is striking how well this value coincides with the point we find in our simulations, i.e. 18.2 Volts. It was also reported that the discharge was very near the point of shut-down, similar to our findings.

The simulations including cx-collisions, which have been shown to be very important in typical dusty plasmas as the one considered here, give a much higher value for the driving potential at void closure. We do not propose that cx-collisions should be ignored altogether, however, close to void-closure they might not be the most important effect. When the dust density inside the void increases, there might be a point where the inter-particle distance becomes of the same order as the Debye length. In this case, the shielding length should no longer be compared to the ion-neutral mean free path, but with the inter-particle distance. Furthermore, OML theory is no longer valid, since this theory is derived for isolated particles, so that a totally different description becomes necessary altogether. Finally, it was recently shown \cite{Hutchinson2007} that the effect of cx-collisions on the ion current to a floating sphere is less than predicted by the generally accepted method we use, discussed in \cite{LampePRL2001}. This means that our solutions with cx-collisions overestimate the additional ion current, however, our best solution to the experiment does not include cx-collisions at all.

The ultimate goal of void-closure is to create large, homogeneous, void-free dust crystals, which can be used to study many phenomena in three dimensions, which otherwise would not be accessible with ordinary optical techniques. Therefore, we need to see whether or not the final dust clouds in the dusty plasmas with closed voids provide such systems.

\section{Dust clouds obtained after void-closure}

\begin{figure}[h]
\includegraphics[width=0.505\textwidth]{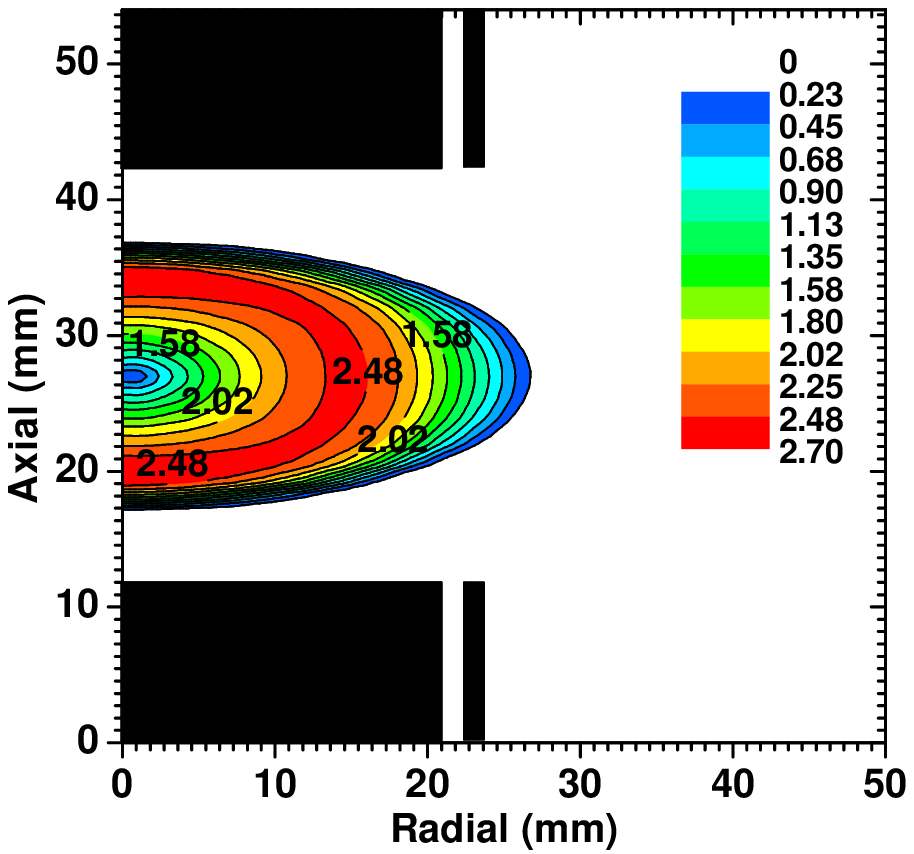}
\includegraphics[width=0.495\textwidth]{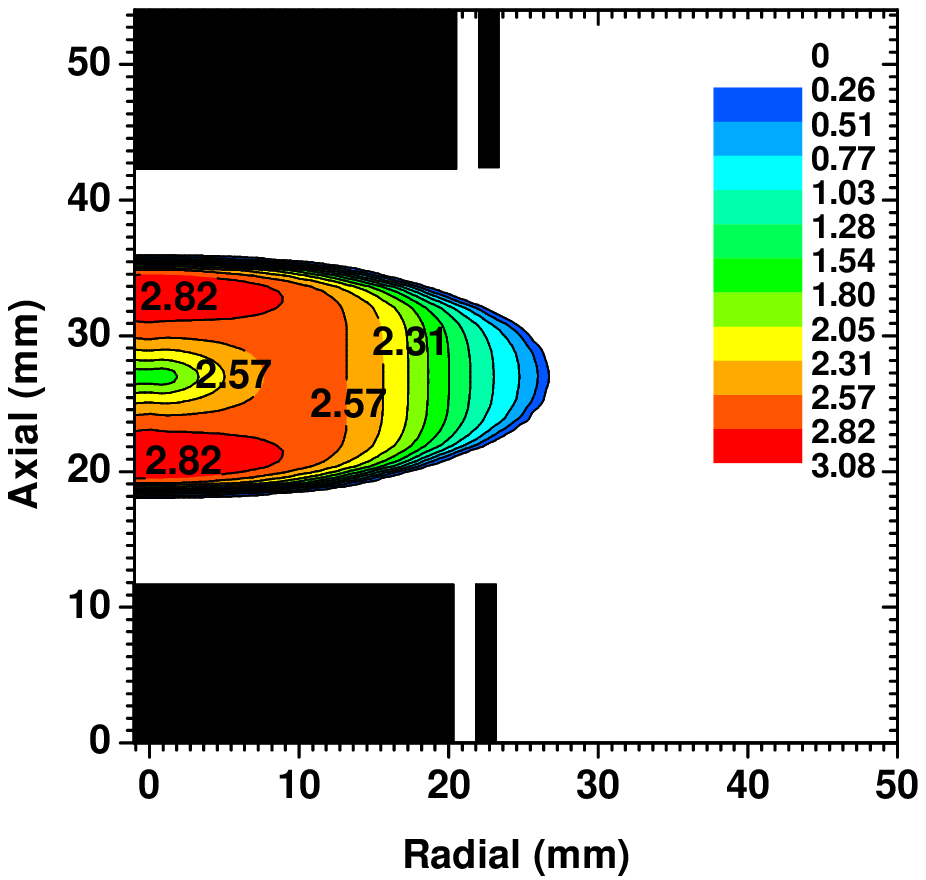}
\caption{\textbf{Left:} The final dusty density distribution (${10}^{-3}$m${}^{-3}$) for 500.000 dust paritcles, without cx-collisions, showing how there still is a strong minimum of the dust density in the center, where the void used to be. \textbf{Right:} The same as on the left, but now with taking cx-collisions into account. The anisotropy is smaller, but is still there.}\label{fig:finaldustclouds500k}
\end{figure} 

\begin{figure}[h]
\includegraphics[width=0.5\textwidth]{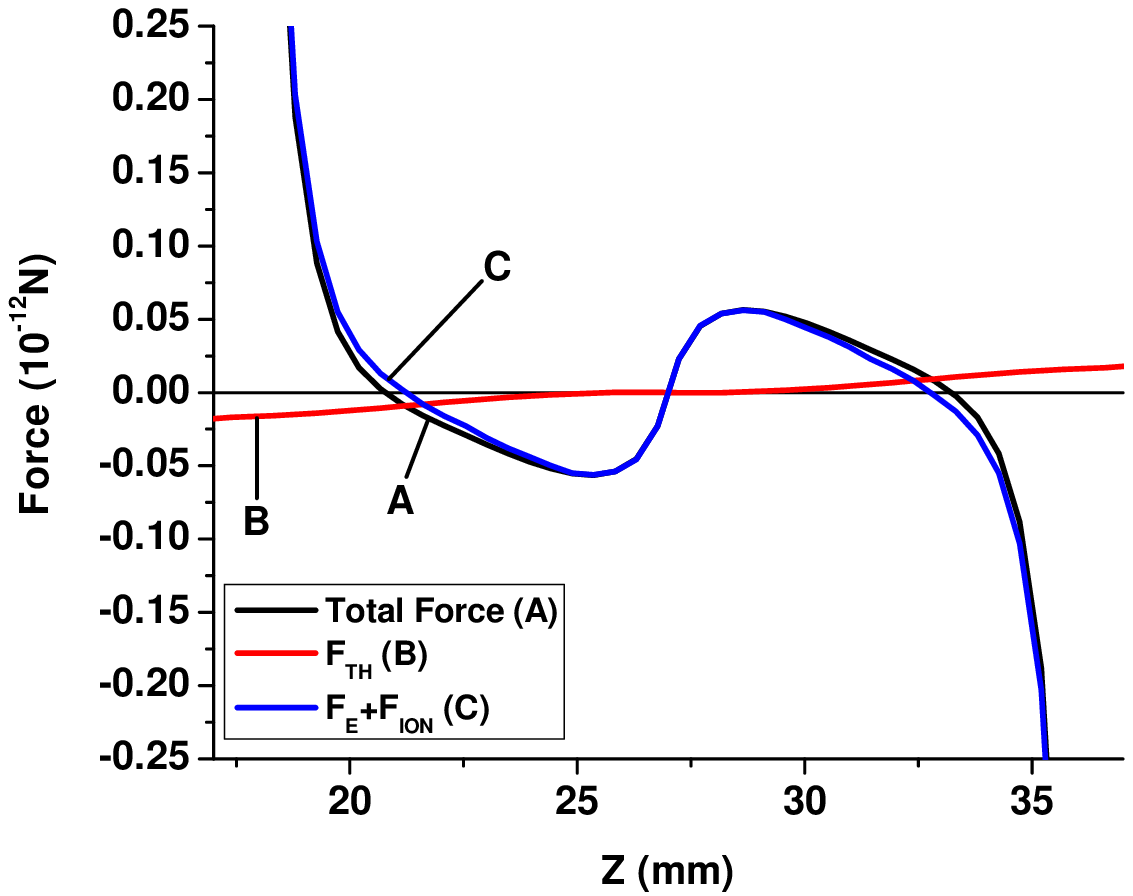}
\includegraphics[width=0.5\textwidth]{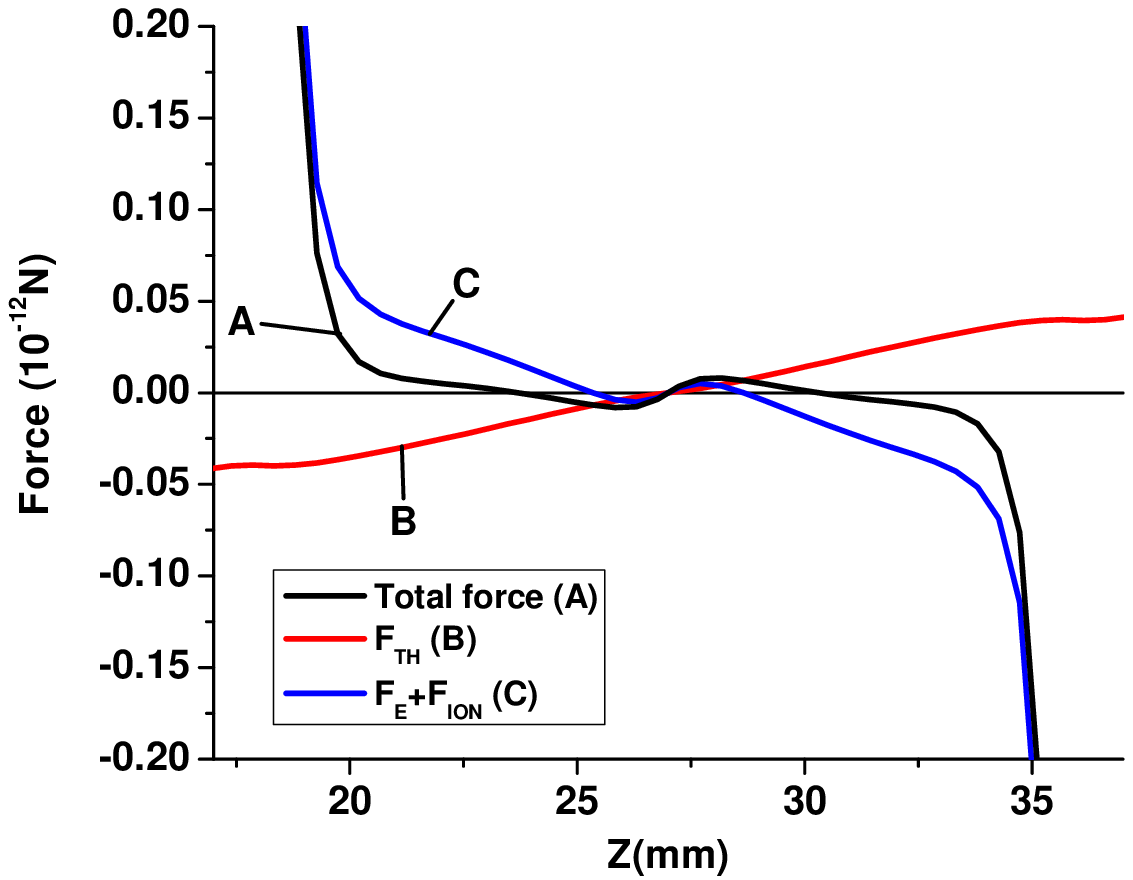}
\caption{\textbf{Left:} The axial forces in the dust cloud for 500.000 added particles, without cx-collisions. There is a residual outward ion drag force, which acts against the internal dust pressure, and causes the dip in dust density. \textbf{Right:} The same as on the left, but now with taking cx-collisions into account. The electrostatic force and ion drag force almost completely vanish. Due to the increased plasma depletion, the background gas is heated more, however, causing an increased thermophoretic force, which enhances the anisotropy.}\label{fig:forces500k}
\end{figure}

First, we show the final dust clouds in the dusty plasma discharges with 500.000 added dust particles. Figure \ref{fig:finaldustclouds500k} shows the solution without collisions on the left, the solution with collisions on the right. It is clear that, even though the void is closed according to our definition, the dust cloud is certainly not homogeneous. The reason for the anisotropy in the dust density is not the same for both solutions. Obviously, there is a small residual force acting against the internal dust pressure, which tries to close the void. When we look at the axial force in the dust cloud, shown in figure \ref{fig:forces500k}, without collisions on the left, with collisions on the right, we see that the anisotropy in the first case is caused by an incomplete cancellation of the outward ion drag force, whereas in the latter case the ion drag and electrostatic force have almost vanished. In this case however, the thermophoretic force is significant, compared to the ion drag and electrostatic force. The thermophoretic force is larger due to the enhanced plasma-recombination on the dust, the result of the enhanced ion current towards the dust. The recombination energy heats the dust particles. Neutral atoms impinging on the dust return back to the gas slightly hotter, causing a gradient in the neutral gas temperature.

\begin{figure}[h]
\includegraphics[width=0.5\textwidth]{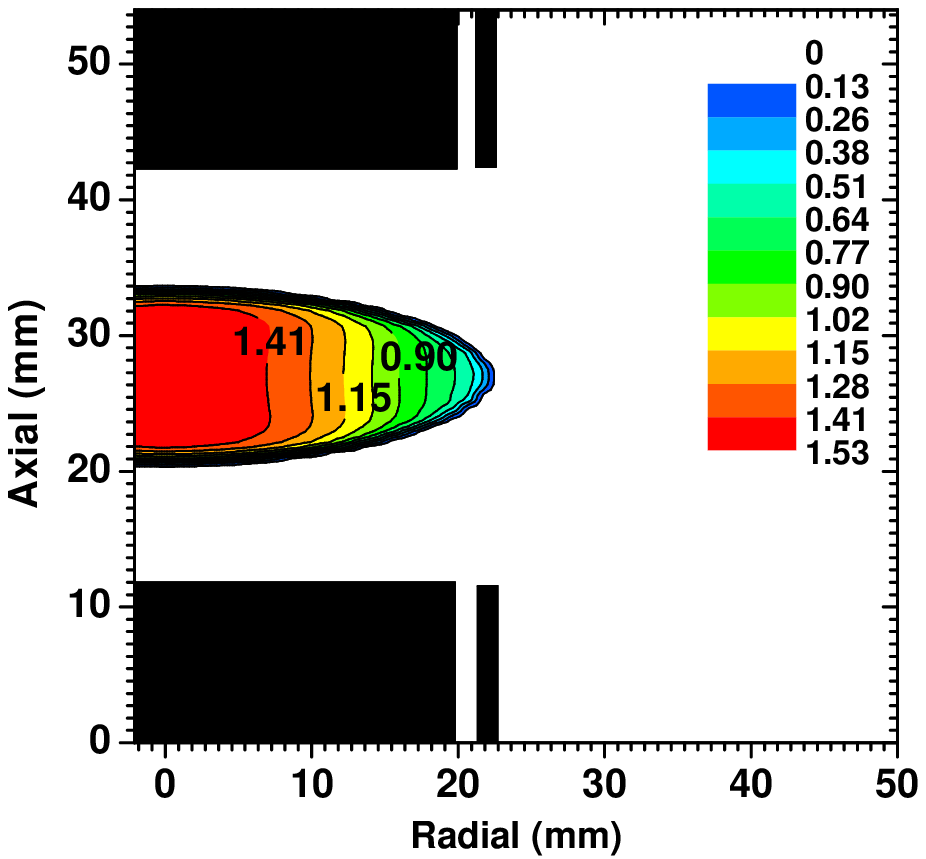}
\includegraphics[width=0.5\textwidth]{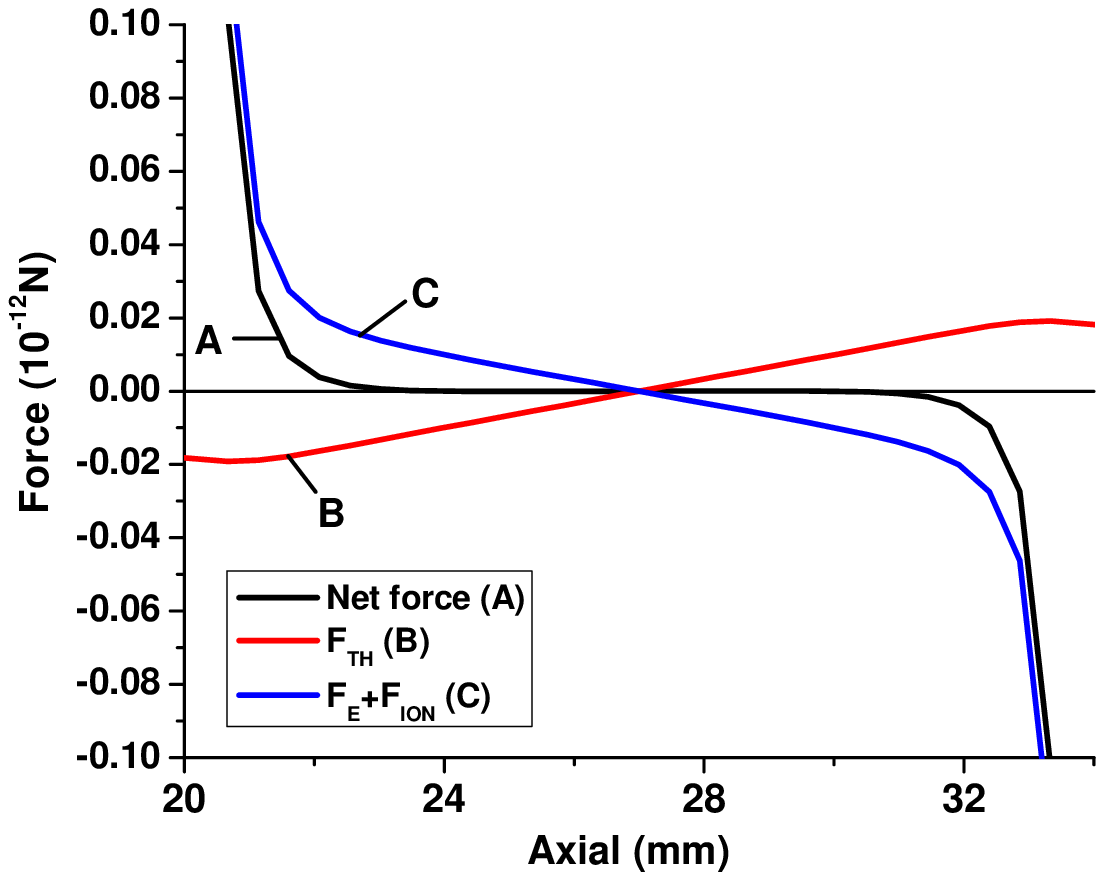}
\caption{\textbf{Left:} The final dust cloud in the discharge with 130.000 particles added and with the effect of cx-collisions on the charging. A large, 3D, homogeneous, dust-free crystal has formed. \textbf{Right:} The forces acting on the dust inside the cloud. The sum of the electrostatic and ion drag force has changed sign, only showing one point where it vanishes, much like an electronegative discharge. The thermophoretic force is small, but still comparable to the other forces, creating a large volume where the dust transport is determined by the internal pressure only.}\label{fig:finaldustcloud130k}
\end{figure}

When we introduce less particles, the effect of the background gas heating must be less, since the plasma depletion on the dust is less. Indeed, the final dust cloud for 130.000 particles, with cx-collisions, shown in figure \ref{fig:finaldustcloud130k} on the left is homogeneous, very suitable for the experiments mentioned before. On the right, the same figure shows the forces throughout the dust cloud. The ion drag force and electrostatic force act void-closing, and have almost vanished, due to the very low space charge.  The dusty plasma in the cloud acts much like an electronegative discharge that way. The thermophoretic force is smaller and almost equal in magnitude, but opposite in sign to these forces. The net effect is a dust cloud where only the dust pressure determines the force, causing a very homogeneous dust cloud.

\section{Conlcusions}

Using a fully self-consistent fluid model for dusty plasmas, we have studied the behavior of dusty plasma under micro-gravity conditions, while closing the void by lowering the driving potential. The simulations without dust show how the virtual void closes with decreasing potential. This is due to the decrease in plasma density, which causes a decrease of the outward ion drag. For decreasing density, and increasing Debye length, the ion drift velocity can increase, but as long as the drift inside the void is subthermal (as it usually is), the density profile determines the shift of the virtual void inwards. The flattening of this profile along the symmetry axis explains why for high driving potential this shift is gradual, but becomes much faster for lower driving potentials. The effect of charge-exchange collisions on the virtual void is a decrease of this void-size for all potentials, since the calculated charge on the dust becomes less.

The real void-size, when many dust particles are added, shows a similar behavior, but now the plasma density behaves completely different, due to the additional depletion of plasma on the dust. The corresponding self-organization of the dusty plasma system results in additional ions that provide the outward force to maintain the void. Once the driving potential becomes too low to maintain this self-organization, the void closes rapidly due to a sudden loss of plasma. This is even more pronounced with charge-exchange collisions, due to the additional plasma losses. When the void closes, essentially a negative dust particle-positive ion plasma is formed, which can be seen by the large difference between the free ion and electron density inside the dust cloud.

The results from void-closure experiments that were done aboard the ISS, in the same reactor, at the same input settings, were well reproduced for simulations with a total of 500.000 particles added, while \emph{not} taking the charge-exchange collisions into account when calculating the dust charge, especially close to the point of void-closure. This might be due to the fact that at that point, the dust density becomes so high that the inter-particle distance becomes comparable to the Debye length. This might reduce the effect of charge-exchange collisions, so that the results without these are closer to the experimental points. Also, an OML-based description of the dust charging is no longer valid, since this theory is derived for isolated particles. A more thorough description would be needed to quantitatively describe the points near void-closure, which apparently are less influenced by cx-collisions than previously expected.

The final dust distributions are found to be homogeneous only when charge-exchange collisions are taken into account and when not too many dust particles are added to the discharge. The electronegative character of the dust cloud plays an important role in minimizing the forces inside the dust cloud, together with the thermophoretic force, due to the enhanced background gas temperature through the additional plasma recombination. This, in a sense, is another example of self-organization in dusty plasma. These dust crystals are so stable, that they allow the dusty plasma to be run at a rather large range of potentials below the point of void closure. This means that these crystals can also be formed by following the opposite scheme, namely by running a plasma at low potential and by adding small batches of dust, while increasing the potential. This way, the discharge is less likely to die out, which might provide a more stable way of forming void-free dust crystals, which are needed for the study of many phenomena on a scale directly visible to the naked eye.

\end{document}